\documentclass[12pt,a4paper]{article}
\usepackage{amsfonts,latexsym}
\usepackage{amsmath,amssymb}
\usepackage{graphicx,color}

\renewcommand{\thefootnote}{\fnsymbol{footnote}}

\begin{document}

\vspace{12mm}

\begin{center}
{{{\Large {\bf Black holes in new massive conformal gravity }}}}\\[10mm]

{Yun Soo Myung$^a$\footnote{e-mail address: ysmyung@inje.ac.kr} and De-Cheng Zou$^{a,b}$\footnote{e-mail address: dczou@yzu.edu.cn}}\\[8mm]

{${}^a$Institute of Basic Sciences and Department  of Computer Simulation, Inje University Gimhae 50834, Korea\\[0pt] }

{${}^b$Center for Gravitation and Cosmology and College of Physical Science and Technology, Yangzhou University, Yangzhou 225009, China\\[0pt]}
\end{center}
\vspace{2mm}

\begin{abstract}
We investigate the black holes in the new massive conformal gravity which is not
invariant under conformal transformations because of the presence of the Einstein-Hilbert term.
First, we show that the small Schwarzschild
black hole is unstable against the $s$-mode of linearized Ricci tensor by solving the  Lichnerowicz-Ricci tensor equation. This instability induces the appearance of the non-BBMB (Bocharova-Bronnikov-Melnikov-Bekenstein) black hole that has both Ricci tensor and conformal scalar hair.
\end{abstract}
\vspace{5mm}

\vspace{1.5cm}

\hspace{11.5cm}{Typeset Using \LaTeX}
\newpage
\renewcommand{\thefootnote}{\arabic{footnote}}
\setcounter{footnote}{0}


\section{Introduction}
Massive conformal gravity (MCG) was introduced  as a model of
the massive gravity  whose  action
consists  of the conformally coupled scalar to the  Einstein-Hilbert term (CCSE)
and the Weyl term~\cite{Faria:2013hxa,Myung:2014aia}.
This action is invariant under conformal transformations. The different actions including the MCG have  been investigated for the other aspects~\cite{Flanagan:1996gw,Mannheim:2001kk,Mannheim:2005bfa,Flanagan:2006ra,Bouchami:2007en,Maldacena:2011mk}.
The MCG might not be a promising model of
the massive gravity~\cite{Myung:2014aia,Myung:2014tna} because one could not obtain the condition of  $\delta R=0$ due to the conformal symmetry.
The non-propagation of the linearized Ricci scalar ($\delta R=0$) is considered as a
strong condition to achieve a massive gravity theory at the
linearized level. In this direction,  the inclusion of  $R$ breaking  conformal symmetry leads to the new massive conformal gravity (NMCG).
Excluding the Weyl term from the NMCG  corresponds  to a famous action which gives us the BBMB black hole with a conformal scalar hair~\cite{Bocharova:1970skc,Bekenstein:1974sf}.
Also, deleting the CCSE in the NMCG leads to the Einstein-Weyl gravity where a single branch for  non-Schwarzschild black holes with Ricci tensor hair was found recently~\cite{Lu:2015cqa}. Here, it is very important to note that the appearance of the non-Schwarzschild black hole is closely related to the instability of Schwarzschild black hole~\cite{Lu:2017kzi,Stelle:2017bdu,Myung:2013doa,Whitt:1985ki}.

On the other hand, very recently, infinite scalarized (charged) black holes were obtained  from the Einstein-Gauss-Bonnet-Scalar theory~\cite{Doneva:2017bvd,Silva:2017uqg} (Einstein-Maxwell-Scalar theory~\cite{Herdeiro:2018wub})
through the coupling of a  scalar to the Gauss-Bonnet term (Maxwell term). In this case, the linearized scalar equation played the important role in determining infinite branches of the $n=0,1,2,\cdots$ scalarized black holes.

In this work, we propose  the NMCG as a promising  candidate for providing  a black hole with Ricci tensor and conformal scalar hairs.
For this purpose,    we wish to  perform
the stability analysis of  the Schwarzschild
black hole without hairs  by making use of  the  Lichnerowicz-Ricci tensor equation (\ref{slin-eq}). Here the linearized scalar equation (\ref{nlsca}) always provides stable modes because its potential is positive definite outside the event horizon.
By solving (\ref{slin-eq}) numerically with time-dependence $e^{\Omega t}$, the small Schwarzschild black hole is unstable
against the $s$-mode of Ricci-tensor perturbation under the
small mass condition of $0<m_2 \le m_{\rm th}=\frac{0.876}{r_+}$ for a massive spin-2 mode. Here $m_{\rm th}$  and $r_+$ denote the threshold mass for instability and  the black hole
horizon size, respectively.  We suggest that the threshold of instability for Schwarzschild black hole determines the appearance of a non-BBMB black hole  with Ricci tensor and conformal scalar hairs. For $m_2=0.845<m_{\rm th}$, we find the non-BBMB black hole solution  by solving three equations (\ref{sbeq-1})-(\ref{sbeq-3}) numerically. Importantly, this numerical solution with primary hair  is different from the BBMB black hole solution with secondary hair.

The organization of our work is as follows. In section 1, we introduce the feature of the NMCG and the BBMB black hole solution obtained in the limit of $m^2_2 \to \infty$.
We review mainly the instability of Schwarzschild black hole without hairs which may imply the appearance of a non-BBMB black hole with Ricci tensor and conformal scalar hairs
in section 3. In section 4 we derive the non-BBMB black hole solution numerically which reduces to the BBMB black hole when taking the limit of $m^2_2 \to \infty$.
We summarize our results in section 5.

\section{New massive conformal gravity}

We begin with the new massive conformal gravity (NMCG) action~\cite{Myung:2014aia,Myung:2014tna}
\begin{eqnarray}S_{\rm NMCG}=\frac{1}{16 \pi G}\int d^4 x\sqrt{-g}
\Big[R-\alpha\Big(\phi^2R+
6\partial_\mu\phi\partial^\mu\phi\Big)-\frac{1}{2m^2_2}C^{\mu\nu\rho\sigma}C_{\mu\nu\rho\sigma}\Big],
\label{NMCG}
\end{eqnarray}
where the second term corresponds to the CCSE with parameter $\alpha$ and the last one is the Weyl term with  mass parameter $m_2^2$.
Hereafter we choose $\alpha=1$ and $G=1$ for simplicity.
Excluding the  Einstein-Hilbert term from (\ref{NMCG}) leads  to the
MCG
which  is invariant under the  conformal transformations
as~\cite{Faria:2013hxa}
\begin{equation} \label{cft}
g_{\mu\nu} \to \Omega^2(x)g_{\mu\nu},~~\phi \to \frac{\phi}{\Omega}.
\end{equation}
Here $\Omega(x)$ is an arbitrary function of the spacetime
coordinates.  We note that adding the Einstein-Hilbert term ($R$) breaks conformal symmetry in
the MCG, leading to the NMCG (\ref{NMCG}). In this case,   the name of NMCG for the action (\ref{NMCG}) might be misleading because conformal symmetry is broken.
However,  we would like to mention that `C' in the NMCG means `conformally coupled scalar to the metric'.

We note that the $R$+CCSE theory appears as a part of the Horndeski gravity which is considered as a general scalar-tensor theory with second-order equations ~\cite{Horndeski:1974wa,Kobayashi:2014wsa,Ganguly:2017ort}. Informing that the Gauss-Bonnet term is a topological surface term in four dimensional spacetime, a general fourth-order gravity is given by
$\gamma R-\alpha C_{\mu\nu\rho\sigma}C^{\mu\nu\rho\sigma}+\beta R^2$~\cite{Lu:2017kzi,Stelle:2017bdu} which shows positive-energy spin-2, negative-energy massive spin-2 with mass $m_2^2=\gamma/2\alpha$ (ghost), and positive-energy massive spin-0 with $m_0^2=\gamma/6\beta$ without tachyons around the Minkowski spacetime. However, the trace no-hair theorem on its Einstein equation simplifies  the numerical analysis of non-Schwarzschild black hole solutions effectively because requiring $R=0$ $(\beta=0)$ could reduce the third-order equation to  the second-order equation. This is the reason  why we included the Weyl term in (\ref{NMCG}) only  by choosing $\beta=0$.
Considering the action (\ref{NMCG}) around  the Minkowski spacetime, we have positive-energy spin-2, negative-energy massive spin-2 with mass $m_2^2$ (ghost), and a positive-energy massless scalar.

The Einstein equation takes the form
\begin{equation} \label{nequa1}
G_{\mu\nu}=
\Big[\phi^2G_{\mu\nu}+g_{\mu\nu}\nabla^2(\phi^2)-\nabla_\mu\nabla_\nu(\phi^2)+6\partial_\mu\phi\partial_\nu\phi-3(\partial\phi)^2g_{\mu\nu}\Big]+\frac{2}{m^2_2}B_{\mu\nu},
\end{equation}
where the Einstein tensor  is given by $G_{\mu\nu}=R_{\mu\nu}-Rg_{\mu\nu}/2$
and the Bach tensor $B_{\mu\nu}$  is defined by
\begin{eqnarray} \label{equa2}
B_{\mu\nu}&=&  \Big(R_{\mu\rho\nu\sigma}R^{\rho\sigma}-\frac{1}{4}
R^{\rho\sigma}R_{\rho\sigma}g_{\mu\nu}\Big)-\frac{1}{3}
R\Big(R_{\mu\nu}-\frac{1}{4} Rg_{\mu\nu}\Big) \nonumber \\
&+&
\frac{1}{2}\Big(\nabla^2R_{\mu\nu}-\frac{1}{6}\nabla^2Rg_{\mu\nu}-\frac{1}{3}\nabla_\mu\nabla_\nu
R\Big).
\end{eqnarray}
We note that its trace is zero  ($B^\mu~_\mu=0$).

On the other hand, the scalar
equation is given by
\begin{equation} \label{ascalar-eq}
\nabla^2\phi-\frac{1}{6}R\phi=0.
\end{equation}
Taking the trace of (\ref{nequa1}) together with (\ref{ascalar-eq}) leads to
\begin{equation} \label{ricciz}
R=0
\end{equation}
which is used to simplify the scalar equation
(\ref{ascalar-eq}) as a massless scalar equation
\begin{equation} \label{tscalar-eq}
\nabla^2\phi=0.
\end{equation}
It is curious to note that the conformally coupled scalar equation (\ref{ascalar-eq}) is transformed to the minimally coupled scalar equation (\ref{tscalar-eq}) in the NMCG.
Importantly, we mention that  `$R=0$' in (\ref{ricciz}) will play the  crucial role in reducing the third-order equation to the second-order equation in obtaining the new non-BBMB black hole solution. This is because it is not easy to solve the higher-order equation more than second-order when imposing  appropriate boundary conditions numerically for a numerical black hole solution.  So, the MCG could not be a candidate  for deriving the non-BBMB black hole solution because of no achievement for $R=0$.

Before we proceed, we  introduce the BBMB solution found in the limit of $m^2_2 \to \infty$ with $\alpha=4\pi/3$ in the NMCG (\ref{NMCG}).
In this case, assuming a spherically symmetric background, the analytic BBMB solution takes the form as~\cite{Bocharova:1970skc,Bekenstein:1974sf}
\begin{equation}
ds^2_{\rm BBMB}=-\Big(1-\frac{M}{r}\Big)^2dt^2+\frac{dr^2}{\Big(1-\frac{M}{r}\Big)^2}+r^2d\Omega^2_2,~~\tilde{\phi}(r)=\pm \sqrt{\frac{3}{4\pi}}\frac{M}{r-M}, \label{bbmb}
\end{equation}
where $M$ is the mass of the black hole. This is a famous black hole solution with Ricci tensor hair and conformal scalar (secondary) hair.  This line element is exactly the same form of  the extremal Reissner-Nordstr\"om black hole with $R_{\mu\nu}\not=0$, but
the conformal scalar hair blows up at the horizon. To obtain  a smooth scalar hair at the horizon,  one might introduce either  the cosmological constant as in the MTZ black hole~\cite{Martinez:2002ru} or the bi scalar tensor theory to find a black hole  in asymptotically flat spacetimes~\cite{Charmousis:2014zaa}.
Also, we note  that  this scalar hair is not primary but secondary because the scalar $\tilde{\phi}(r)$ depends on the black hole mass $M$ which is regarded as the only parameter of the solution, and it does not carry an independent scalar charge. Furthermore, the BBMB solution does not have a continuous limit to the Schwarzschild black hole, showing a feature of conformally coupled scalar to the metric.
However, adding the Weyl term to the $R$+CCSE theory
will reveal a non-BBMB solution with two charges of ADM mass $M$ and scalar charge $Q_s$. In this case, the conformal scalar hair becomes primary. We emphasize that the Weyl term makes a shift from the BBMB solution to the non-BBMB solution.

\section{Instability of small Schwarzschild black hole}
First of all, we would like to mention that what follows is mainly  a review of the known results.
Introducing the background ansatz without Ricci-tensor and scalar hairs
\begin{equation} \label{back-val}
\bar{R}_{\mu\rho\nu\sigma}\not=0,~~\bar{R}_{\mu\nu}=0,~~\bar{R}=0,~~\bar{\phi}=0,
\end{equation}
Eqs.(\ref{nequa1}) and (\ref{tscalar-eq}) imply   the
Schwarzschild black hole solution \begin{equation} \label{schw}
ds^2_{\rm Sch}=\bar{g}_{\mu\nu}dx^\mu
dx^\nu=-f(r)dt^2+\frac{dr^2}{f(r)}+r^2d\Omega^2_2
\end{equation}
with the metric function \begin{equation} \label{num}
f(r)=1-\frac{r_+}{r}.
\end{equation}
The event horizon is located at $r=r_+=2M$. It is worth noting that  the Schwarzschild black hole is also a solution to the  $R+$CCSE theory
which admits the BBMB solution.

Let us consider the metric and scalar perturbations around the
Schwarzschild  black hole
\begin{eqnarray} \label{m-p}
g_{\mu\nu}=\bar{g}_{\mu\nu}+h_{\mu\nu},~~\phi=0+\varphi.
\end{eqnarray}
Linearizing (\ref{nequa1}) around (\ref{schw}) with (\ref{back-val})  leads to the linearized Einstein equation
\begin{eqnarray} \label{nlin-eq}
 m^2_2\delta G_{\mu\nu}=\bar{\nabla}^2\delta
G_{\mu\nu}+2\bar{R}_{\rho\mu\sigma\nu}\delta G^{\rho\sigma}
-\frac{1}{3}\Big(\bar{\nabla}_\mu\bar{\nabla}_\nu-\bar{g}_{\mu\nu}\bar{\nabla}^2
\Big) \delta R,
\end{eqnarray}
which is completely decoupled from the scalar perturbation $\varphi$ because of $\bar{\phi}=0$.
For the Schwarzschild black hole with constant scalar hair and its linearization, see Ref.~\cite{Myung:2014tna}.
Here, the linearized Einstein tensor, Ricci tensor, and Ricci scalar are given by
\begin{eqnarray}
\delta G_{\mu\nu}&=&\delta R_{\mu\nu}-\frac{1}{2} \delta
R\bar{g}_{\mu\nu},
\label{ein-t} \\
\delta
R_{\mu\nu}&=&\frac{1}{2}\Big(\bar{\nabla}^{\rho}\bar{\nabla}_{\mu}h_{\nu\rho}+
\bar{\nabla}^{\rho}\bar{\nabla}_{\nu}h_{\mu\rho}-\bar{\nabla}^2h_{\mu\nu}-\bar{\nabla}_{\mu}
\bar{\nabla}_{\nu}h\Big), \label{ricc-t} \\
\delta R&=& \bar{g}^{\mu\nu}\delta R_{\mu\nu}= \bar{\nabla}^\mu
\bar{\nabla}^\nu h_{\mu\nu}-\bar{\nabla}^2 h. \label{Ricc-s}
\end{eqnarray}
On the other hand, taking into account (\ref{tscalar-eq}) leads to the linearized scalar equation
\begin{equation}\label{nlsca}
\bar{\nabla}^2\varphi=0
\end{equation}
whose scalar is a propagating wave being free from unstable
modes~\cite{Myung:2014nua}. Taking the trace of (\ref{nlin-eq})  together with (\ref{nlsca}) implies the non-propagation of linearized Ricci scalar
\begin{equation}
\delta R=0. \label{lris}
\end{equation}
It is noted  that (\ref{lris})  is confirmed by
linearizing $R=0$ (\ref{ricciz}) directly. In  case of  the MCG, one could not  obtain $\delta R=0$.
That is, if one does not break conformal symmetry, one could not achieve
the non-propagation of the linearized Ricci scalar.   Substituting
$\delta R=0$  into Eq. (\ref{nlin-eq}) leads to the
Lichnerowicz-Ricci tensor equation  for the linearized Ricci tensor
\begin{equation} \label{slin-eq}
\bar{\nabla}^2\delta R_{\mu\nu}+ 2\bar{R}_{\rho\mu\sigma\nu}\delta
R^{\rho\sigma}=m^2_2\delta
R_{\mu\nu}\to (\Delta_{\rm L}+m^2_2)\delta R_{\mu\nu}=0,
\end{equation}
where $\Delta_{\rm L}$ is called  the Lichnerowicz operator based on the Schwarzschild black hole background.
We note that Eq.(\ref{slin-eq})  firstly appeared in Ref.~\cite{Whitt:1985ki}, but it was solved for the stability of Schwarzschild black hole correctly in Ref.~\cite{Myung:2013doa}.
We mention that the same expression was remarked in~\cite{Lu:2017kzi,Stelle:2017bdu} when explaining the connection between instability of Schwarzschild black hole and appearance of non-Schwarzschild black hole in the Einstein-Weyl gravity.

It is worth noting  that Eq. (\ref{slin-eq}) may  describe
a massive spin-2 field (5 DOF) propagating around the
Schwarzschild black hole. This is so because $\delta R_{\mu\nu}$
satisfies the transverse and traceless (TT)  condition
\begin{eqnarray}\label{R-TT}
\bar\nabla^{\mu}\delta R_{\mu\nu} =0,~\delta R=0,
\end{eqnarray}
where the contracted Bianchi identity  was used to prove the
transverse condition. Furthermore, if one introduces the TT gauge for metric  perturbation
\begin{eqnarray} \label{h-TT}
\bar\nabla^{\mu}h_{\mu\nu} =0,~h=0,
\end{eqnarray}
one finds the relation $\delta R=\frac{1}{2}\Delta_{\rm L} h_{\mu\nu}$.
In this case, the Lichnerowicz equation (\ref{slin-eq}) leads to the fourth-order equation  as
\begin{equation}
\Big(\Delta_{\rm L}+m^2_2\Big)\Delta_{\rm L} h_{\mu\nu}=0.
\end{equation}
Its general solution  is given by a linear  superposition of solutions to $\Delta_{\rm L} h_{\mu\nu}=0$ and the Lichnerowicz equation for  $h_{\mu\nu}$ as
\begin{equation}
\Big(\Delta_{\rm L}+m^2_2\Big) h_{\mu\nu}=0,
\end{equation}
which is the same equation  obtained for the $s$-mode of metric
perturbation around the Schwarzschild black hole in the
dRGT gravity~\cite{Babichev:2013una,Brito:2013wya}.

In order to understand the origin of the Lichnerowicz equation,  we may introduce
a famous model of the  five-dimensional black string (BS)~\cite{Gregory:1993vy}
\begin{equation}
ds^2_{\rm BS}=ds^2_{\rm Sch}+dz^2.
\end{equation}
We have perturbation along an extra direction of the $z$-axis
\begin{eqnarray}
h_{MN}(t,{\bf x},z)=e^{ikz}e^{\tilde{\Omega} t}\left(
\begin{array}{cc}
h_{\mu\nu}({\bf x}) & 0 \cr 0 & 0
\end{array}
\right). \label{evenpt}
\end{eqnarray}
Using the TT gauge condition (\ref{h-TT}), the linearized equation $\delta R_{MN}=0$ reduces to
\begin{equation} \label{hmn-eq}
\bar{\nabla}^2h_{\mu\nu}+2\bar{R}_{\rho\mu\sigma\nu}h^{\rho\sigma}=k^2h_{\mu\nu} \to \Big(\Delta_{\rm L}+k^2\Big) h_{\mu\nu}=0
\end{equation}
which describes a massive spin-2 mode with 5 DOF propagating
around the Schwarzschild black hole. One has found a long wavelength
perturbation of $0<k<k_c= \frac{{\cal O}(1)}{r_+}$ along  $z$-axis, which
induces  an unstable mode of $e^{\tilde{\Omega}t}$. This picture  indicates  the
Gregory-Laflamme (GL) instability in the BS theory.
If one compares
(\ref{slin-eq}) with (\ref{hmn-eq}), they are exactly the
same when replacing $\delta R_{\mu\nu}$ and $m^2_2$ by $h_{\mu\nu}$ and
$k^2$.   This implies that the instability of the Schwarzschild black hole in the
NMCG arose from the massiveness of $m^2_2\not=0$ where the geometry
of extra $z$ dimension trades for the  mass.

Let us  solve the Lichnerowicz equation (\ref{slin-eq}) with $\delta R_{\mu\nu}(t,{\bf x})=e^{\Omega t} \delta  \tilde{R}_{\mu\nu}({\bf x})$.
 The $s(l=0)$-mode in polar sector of Eq.(\ref{slin-eq}) satisfies the Schr\"{o}dinger-type quation with a tortoise coordinate $r_*=\int [dr/f(r)]$~\cite{Brito:2013wya,Lu:2017kzi}
      \begin{equation} \label{EM-s}
\frac{d^2\delta \tilde{R}^{l=0}_{\mu\nu}}{dr_*^2}-\Big[\Omega^2+V_Z(r)\Big]\delta\tilde{ R}^{l=0}_{\mu\nu}=0
\end{equation}
where $\delta \tilde{R}^{l=0}_{\mu\nu}$ is an $s$-mode of $\delta\tilde{ R}_{\mu\nu}$ and the Zerilli potential $V_Z(r)$ is given by
\begin{equation} \label{EW-p}
V_Z(r)=f(r)\Bigg[\frac{r_+}{r^3}+m^2_2-\frac{12r_+(r-0.5r_+)m^2_2+6r^3(2r_+-r)m^4_2}{(r_++r^3m^2_2)^2}\Bigg].
\end{equation}
\begin{figure*}[t!]
   \centering
   \includegraphics{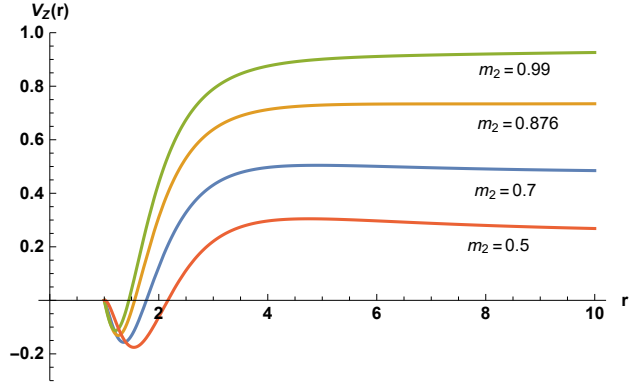}
\caption{Shapes of Zerilli potential $V_Z(r)$ as function of $r\in[r_+=1,10]$ with four different mass $m_2=0.99~({\rm stable}),0.876~({\rm threshold}),0.7~({\rm unstable}),
0.5~({\rm unstable})$. All potentials never  approach zero  asymptotically  because of non-zero  mass term $m_2^2$.  }
\end{figure*}
All potentials with $m_2<m_{\rm th}$ develop negative region near the horizon, whereas their asymptotic limits are positive constants ($V_Z \to m^2_2$, $r\to \infty$).
As is shown in Fig. 1, the negative region becomes wide and deep as the  mass decreases, implying the instability of Schwarzschild black hole.
Solving Eq.(\ref{EM-s}) with appropriate boundary conditions numerically, one finds unstable tensor modes from Fig. 2.
\begin{figure*}[t!]
   \centering
   \includegraphics{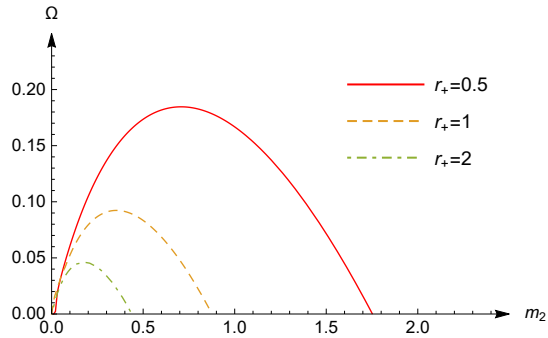}
\caption{Picture for unstable $s$-modes of Ricci tensor perturbation with three different horizon radii based on the Schwarzschild black hole.
The $y[x]$-axis represent $\Omega$ in $e^{\Omega t}$ [mass $m_2$ of massive spin-2 mode].  We read off the thresholds of instability ($\Omega=0$ for $m_2\not=0$):  $m_{\rm th}=$1.752, 0.876, 0.438 for $r_+=0.5,~1,~2$.}
\end{figure*}
From Fig. 2, the GL instability mass bound for $s(l=0)$-mode is given by
\begin{equation} \label{GL-bound}
0<m_2<m_{\rm th}=\frac{0.876}{r_+},
\end{equation}
where $m_{\rm th}$ represents the threshold of GL instability: $m_{\rm th}=1.752,0.876,0.438$ for $r_+=0.5,1,2$.
Here, choosing $m_{\rm th}=0.876$ for $r_+=r_{\rm c}=1$, we obtain the bound for unstable small black holes with horizon radius
\begin{equation}
r_+<r_{\rm c}.
\end{equation}
Also, $m_2=m_{\rm th}=0.876$ denotes  a bifurcation point which allows a  non-BBMB black hole with Ricci tensor and scalar hairs.
Any choice $m_2 \in (0,m_{\rm th}]$ in the single branch will provide a non-BBMB black hole with Ricci tensor and scalar hairs in the next section.

Consequently, the instability of Schwarzschild black hole without hairs may imply the appearance of a non-BBMB black hole with Ricci tensor and conformal scalar hairs. These black hole solutions do not have to be infinitesimally close to each other except at  bifurcation point where
different branches of solutions coalesce.

\section{Non-BBMB black holes}

In order to find a new  non-BBMB black hole, we introduce a spherically symmetric metric
\begin{equation} \label{non-bbmb}
ds^2_{\rm non-BBMB}=g_{\mu\nu}dx^\mu
dx^\nu=-A(r)dt^2+\frac{dr^2}{B(r)}+r^2d\Omega^2_2
\end{equation}
with conformal  scalar $\phi(r)$. Comparing with the BBMB  case (\ref{bbmb}), we prefer to choose  $A(r)\not= B(r)$ for   a numerical solution.

Before we proceed, we mention that the BBMB black hole with $A=B$ is the unique static, asymptotically flat solution of the $R$+CCSE theory~\cite{XZ}.
A conformally coupled scalar on the solution geometry plays the role of  an electric charge at extremality.
However, one knows   an open problem which could be resolved numerically determining the exact nature of  this solution~\cite{Charmousis:2015aya}.
In this work, we will obtain  the BBMB black hole numerically, starting  from the  $R$+CCSE theory and (\ref{non-bbmb}).

Plugging (\ref{non-bbmb}) with $\phi(r)$  into (\ref{nequa1}) and (\ref{tscalar-eq}), one finds three equations initially: $(t,t)\to$ fourth-order equation; $(r,r)\to$ third-order equation; $(\theta,\theta)\to$ fourth-order equation with   second-order equation for scalar. A reducing mechanism of third-order to second-order is performed by
using ($r,r$)-equation together with $R=0$ to arrive at  a reduced ($r,r$)-equation.
The three second-order equations of $R=0,\nabla^2 \phi=0$, and the reduced ($r,r$)-equation are given by
\begin{eqnarray}
&& \frac{r^2BA'^2-4A^2(-1+B+rB')-rA[rA'B'+2B(2A'+rA'')]}{2r^2A^2}=0, \label{sbeq-1} \\
&& \phi''+\Big(\frac{2}{r}+\frac{B'}{2B}+\frac{A'}{2A}\Big)\phi'=0,\label{sbeq-2} \\
&&(-2A+rA')B''+\frac{[4A^2(-1+B)+2rAA'B+r^2A'^2B]B'}{2rAB}+\frac{3AB'^2}{2B}  \nonumber \\
&&+\frac{3r^2AA'^2B^2-r^3A'^3B^2+2m^2_2[2r^3A^2A'B+(-1+\phi^2+2\phi\phi')]}{2r^2A^2B} \label{sbeq-3} \\
&&+\frac{2m^2_2A\Big(\frac{2B^2}{m^2_2}+r^2(1-\phi^2)+B[\frac{2}{m^2_2}-r^2+r^2\phi^2+4r^3\phi\phi'+3r^4\phi'^2]\Big)}{r^2B}=0. \nonumber
\end{eqnarray}
Considering the event horizon at $r=r_+$, one suggests an approximate solution $[A(r),B(r),$ $\psi(r)=1/\phi(r)$] to the above equations in the near-horizon limit
\begin{equation} \label{nearHs}
A(r)=\sum_{i=2}^\infty A_i(r-r_+)^i,~~B(r)=\sum_{i=2}^\infty B_i(r-r_+)^i,~~\psi(r)=\sum_{i=1}^\infty\psi_i(r-r_+)^i,
\end{equation}
where the first four coefficients are given by
\begin{eqnarray}
&&\underline{A_2},~A_3=\frac{2(-4+r_+^2\psi_1^2)A_2}{3r_+},~A_4=\frac{(41-19r_+^2 \psi_1^2+5r_+^4 \psi^4_1)A_2}{9r_+^2},\nonumber \\
&& A_5=\frac{[-874-267r_+^4 \psi_1^4+70r_+^6\psi_1^6+9r_+^2\psi_1^2(59-\frac{18\psi^2_1}{m^2_2})]A_2}{135r_+^3}, \nonumber \\
&&B_2=\frac{1}{r_+^2},~B_3=-\frac{2(2+r_+^2\psi_1^2)}{3r_+^3},~B_4=\frac{5+5r_+^2\psi_1^2-r_+^4\psi_1^4}{3r_+^4}, \nonumber \\
&&B_5=\frac{-86+27r_+^4\psi_1^4-10r_+^6\psi_1^6-3r_+^2\psi_1^2(37-\frac{54\psi_1^2}{m^2_2})}{45r_+^5},   \label{coeffi} \\
&&\underline{\psi_1},~\psi_2=\frac{(-1+r_+^2\psi_1^2)\psi_1}{3r_+},~~\psi_3=\frac{2(-1+r_+^2\psi_1^2)^2\psi_1}{9r_+^2}, \nonumber \\
&& \psi_4=\frac{[-44-117 r_+^4\psi_1^4+50r_+^6\psi_1^6+3r_+^2\psi_1^2(37-\frac{54\psi_1^2}{m_2^2})]\psi_1}{270r_+^3},\nonumber
\end{eqnarray}
where $A_5~B_5,$ and $\psi_4$ include the mass term $m_2^2$. Here, we introduce  $\psi=1/\phi$-expansion because $\phi$ blows up at the horizon.

In case of $m_2^2\to \infty$, we find  relations $\psi_1=1/r_+$ and $A_2=1/r_+^2$ from Table 1. We recover the series form for the BBMB solution (\ref{bbmb}) in the near-horizon limit
\begin{eqnarray}
 A(r)&=&B(r)=\frac{(r-r_+)^2}{r_+^2}-\frac{2(r-r_+)^3}{r_+^3}+\frac{3(r-r_+)^4}{r_+^4}-\frac{4(r-r_+)^5}{r_+^5}+\cdots, \nonumber \\
  &\to& \Big[\Big(1-\frac{r_+}{r}\Big)^2\Big]_{r=r_+}, \label{series-nh}\\
 \psi(r)&=&\frac{(r-r_+)}{r_+}.\nonumber
\end{eqnarray}
Here we choose  $\alpha=1$ for scalar normalization, compared to the BBMB case of $\alpha=4\pi/3$.
 So, it is worth noting  that the new solution with hairs is based on the BBMB solution  not the Schwarzschild solution (\ref{schw}).
In our theory of the NMCG, two free parameters $A_2$ and $\psi_1$ will be determined  when matching (\ref{nearHs}) with the following asymptotic solution for  $r\gg r_+$:
\begin{eqnarray}
&&A(r)=1-\frac{2M}{r}+\frac{Q_s^2}{r^2}+\frac{Q_s^2(-M^2+Q_s^2+\frac{6}{m^2_2})}{3r^4}+\cdots,\nonumber \\
&&B(r)=1-\frac{2M}{r}+\frac{Q_s^2}{r^2}+\frac{2Q_s^2(-M^2+Q_s^2+\frac{6}{m^2_2})}{3r^4}+\cdots, \label{in-ss} \\
&&\phi(r)=\frac{Q_s}{r}+\frac{M Q_s}{r^2}+\frac{Q_s(4M^2-Q_s^2)}{3r^3}+\frac{MQ_s(2M^2-Q_s^2)}{r^4}+\cdots. \nonumber
\end{eqnarray}
In  case of  $m_2^2\to \infty$, we obtain a relation $M=Q_s$ from Table 1.  One finds  the BBMB solution (\ref{bbmb}) for  $r\gg M$ as
\begin{equation}
A(r)=B(r)=\Big(1-\frac{M}{r}\Big)^2,~~\phi(r)=\frac{M}{r}+\frac{M^2}{r^2}+\frac{M^3}{r^3}+\frac{M^4}{r^4}+\cdots=\Big[\frac{M}{r-M}\Big]_{r\gg M}. \label{bbmbs}
\end{equation}
\begin{figure*}[t!]
   \centering
   \includegraphics{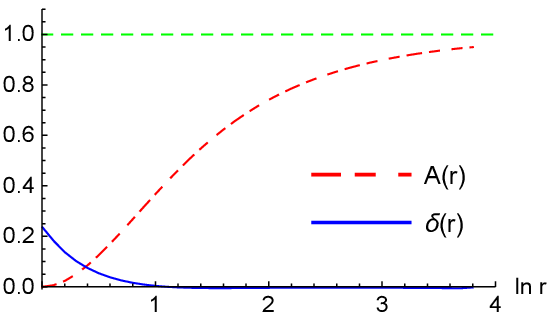}
   \hfill%
   \includegraphics{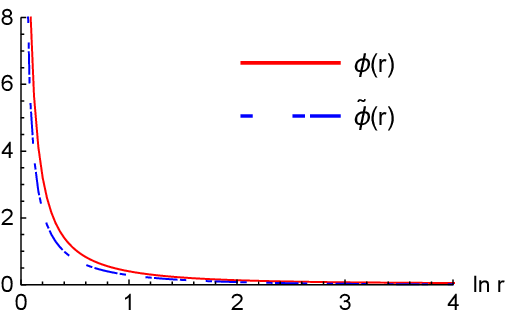}
\caption{Plots of a non-BBMB black hole with $m_2=0.845$ and $Q_s=0.55$ in the single branch of $ m_2\in(0,m_{\rm th}]$  in the NMCG. (Left) Metric function $A(r)$ and $\delta(r)=\ln[B/A]/2$ and the green dotted line represents asymptotic limit of $A(r) \to 1$. For  the BBMB solution, one finds  $\delta(r)=0$ because of $A=B$. (Right) Scalar hairs for non-BBMB  $\phi(r)$ and  BBMB $\tilde{\phi}(r)$ in (\ref{bbmb}) for $\ln r\in[0,4]$. }
\end{figure*}
\begin{table}[h]
\resizebox{7cm}{!}
{\begin{tabular}{|c|c|c|c|c|}
  \hline
$r_+$ & $\psi_1$ & $A_2$ & $M$ & $Q_s$   \\ \hline
0.5&2.000 & 4.000 & 0.5000  &0.5000  \\ \hline
0.75&1.3333 & 1.7778 & 0.7500  &0.7500  \\ \hline
1 &1.000& 1.004 & 1.0002 & 0.9999  \\ \hline
1.25 &0.8000& 0.6400 & 1.2504 & 1.2500  \\ \hline
1.5 &0.6667& 0.4445 & 1.5000 & 1.5008  \\ \hline
\end{tabular}}
\caption{BBMB solution for  different $r_+$ appears in the limit of $m^2_2\to \infty$.
Here, we read off  relations  $\psi_1=1/r_+$ and $A_2=1/r_+^2$ in the near-horizon limit and  $M=Q_s$ for $r\gg r_+$.  }
\end{table}

At this stage, it is emphasized  that the BBMB solution is obtained  numerically from the $R$+CCSE theory ($m^2_2\to \infty$ limit of the NMCG theory).

Any choice of mass $m_2$ less than $m_{\rm th}$ belonging to the single branch is allowed  for getting any non-BBMB solution because it gives negative potential outside the horizon as shown in (\ref{EW-p}) and Fig. 1, implying the instability of Schwarzschild black hole.
Choosing  $r_+=M=1$ and $m_2=0.845<m_{\rm th}=0.876$, we obtain the non-BBMB black hole solution with $A_2=0.635$,$\psi_1=1.335$, and $Q_s=0.55$  which are different from those in the BBMB (See Fig. 3). A choice of $m_2=0.845$  is nothing special and it is introduced for computation.

We wish to call this numerical black hole as the non-BBMB black hole solution by focusing on the role of Weyl term. This is because
the presence of Weyl term in the Einstein-Weyl gravity shifts from the (analytic) Schwarzschild black hole to the (numerical) `non'-Schwarzschild black hole~\cite{Lu:2015cqa}. Here the BBMB black hole is an analytic  solution to the NMCG without  Weyl term, while the `non'-BBMB black hole is a numerical solution to the NCMG (\ref{NMCG}). It is clear that  the Weyl term shifts from the BBMB solution to the non-BBMB solution.
In the non-BBMB solution, the conformal scalar hair is primary because  the scalar charge is  given  independently by  $Q_s=0.55$, compared to the secondary scalar hair ($Q_s=M$) in the BBMB solution.
Any numerical solutions are allowable  for any choice of $m_2\in (0,m_{\rm th}]$ in the single branch.
We find from the non-BBMB solution that $R_{\mu\nu}\not=0$ and $\phi(r)\not=0$, implying the Ricci tensor and conformal  scalar hairs. However, the conformal scalar blows up at the horizon, as in the BBMB solution.

\section{Discussions}
It is well known that the $R+$CCSE theory implies the famous BBMB solution with conformal  scalar hair,
whereas the Einstein-Weyl gravity has indicated  the non-Schwarzschild solution with the Ricci tensor hair. Even though the first corresponds to an analytic solution and the latter is a numerical solution, the condition of $R=0$ played the important  role in deriving
the  black hole solutions in both theories. In this work, we have shown that  the BBMB solution is obtained  numerically from the $R$+CCSE theory.

In the NMCG theory of $R$+CCSE+Weyl term, we have  firstly  performed
the stability analysis of  the Schwarzschild
black hole without hair  by making use of  the  Lichnerowicz-Ricci tensor equation because the linearized scalar equation  always provides stable modes. The small Schwarzschild black hole is unstable
against the $s$-mode of Ricci tensor perturbation under the
condition of $0<m_2 \le \frac{0.876}{r_+}$.

The instability of Schwarzschild black hole without hair determines the appearance of the non-BBMB black hole with Ricci tensor and conformal scalar hairs.
For a choice of  $m_2=0.845 \in (0,m_{\rm th}=0.876]$,  we have found the non-BBMB black hole with Ricci tensor and conformal scalar hairs numerically.
We note that any numerical solutions are available for any  $m_2\in (0,m_{\rm th}]$ in the single branch.
In deriving the solution, the condition of $R=0$ played the important role in reducing the third-order equation to the second-order one.
We wish to mention that  the non-BBMB solution carries two charges of ADM  mass $M$ and scalar charge $Q_s$, and the conformal scalar hair  is primary. However, the conformal scalar hair  blows up at the horizon, like the BBMB solution. Thus, in order to cure it,  one may introduce either  the cosmological constant to obtain a regular scalar hair as in the MTZ black hole~\cite{Martinez:2002ru} or the bi scalar tensor theory to find a black hole with a regular  scalar hair in asymptotically flat spacetimes~\cite{Charmousis:2014zaa}.

 \vspace{2cm}

{\bf Acknowledgments}
 \vspace{1cm}

This work was supported by the National Research Foundation of Korea (NRF) grant funded by the Korea government (MOE)
 (No. NRF-2017R1A2B4002057).


\newpage

\begin{thebibliography}{99}
\bibitem{Faria:2013hxa}
  F.~F.~Faria,
  Adv.\ High Energy Phys.\  {\bf 2014}, 520259 (2014)
  doi:10.1155/2014/520259
  [arXiv:1312.5553 [gr-qc]].

\bibitem{Myung:2014aia}
  Y.~S.~Myung,
  Phys.\ Lett.\ B {\bf 730}, 130 (2014)
  doi:10.1016/j.physletb.2014.01.044
  [arXiv:1401.1890 [gr-qc]].

\bibitem{Myung:2014tna}
  Y.~S.~Myung,
  arXiv:1402.2006 [gr-qc].

\bibitem{Flanagan:1996gw}
  E.~E.~Flanagan and R.~M.~Wald,
  Phys.\ Rev.\ D {\bf 54}, 6233 (1996)
  doi:10.1103/PhysRevD.54.6233
  [gr-qc/9602052].


\bibitem{Mannheim:2001kk}
  P.~D.~Mannheim,
  Int.\ J.\ Mod.\ Phys.\ D {\bf 12}, 893 (2003)
  doi:10.1142/S0218271803003414
  [astro-ph/0104022].


\bibitem{Mannheim:2005bfa}
  P.~D.~Mannheim,
  Prog.\ Part.\ Nucl.\ Phys.\  {\bf 56}, 340 (2006)
  doi:10.1016/j.ppnp.2005.08.001
  [astro-ph/0505266].


\bibitem{Flanagan:2006ra}
  E.~E.~Flanagan,
  Phys.\ Rev.\ D {\bf 74}, 023002 (2006)
  doi:10.1103/PhysRevD.74.023002
  [astro-ph/0605504].


\bibitem{Bouchami:2007en}
  J.~Bouchami and M.~B.~Paranjape,
  Phys.\ Rev.\ D {\bf 78}, 044022 (2008)
  doi:10.1103/PhysRevD.78.044022
  [arXiv:0710.5402 [hep-th]].

\bibitem{Maldacena:2011mk}
  J.~Maldacena,
  arXiv:1105.5632 [hep-th].


\bibitem{Bocharova:1970skc}
  N.~M.~Bocharova, K.~A.~Bronnikov and V.~N.~Melnikov,
  Vestn.\ Mosk.\ Univ.\ Ser.\ III Fiz.\ Astron.\ , no. 6, 706 (1970).

\bibitem{Bekenstein:1974sf}
  J.~D.~Bekenstein,
  Annals Phys.\  {\bf 82}, 535 (1974).
  doi:10.1016/0003-4916(74)90124-9


\bibitem{Lu:2015cqa}
  H.~Lu, A.~Perkins, C.~N.~Pope and K.~S.~Stelle,
  Phys.\ Rev.\ Lett.\  {\bf 114}, no. 17, 171601 (2015)
  doi:10.1103/PhysRevLett.114.171601
  [arXiv:1502.01028 [hep-th]].




\bibitem{Lu:2017kzi}
  H.~Lü, A.~Perkins, C.~N.~Pope and K.~S.~Stelle,
  Phys.\ Rev.\ D {\bf 96}, no. 4, 046006 (2017)
  doi:10.1103/PhysRevD.96.046006
  [arXiv:1704.05493 [hep-th]].


\bibitem{Stelle:2017bdu}
  K.~S.~Stelle,
  Int.\ J.\ Mod.\ Phys.\ A {\bf 32}, no. 09, 1741012 (2017).
  doi:10.1142/S0217751X17410123






\bibitem{Myung:2013doa}
  Y.~S.~Myung,
  Phys.\ Rev.\ D {\bf 88}, no. 2, 024039 (2013)
  doi:10.1103/PhysRevD.88.024039
  [arXiv:1306.3725 [gr-qc]].

\bibitem{Whitt:1985ki}
  B.~Whitt,
  Phys.\ Rev.\ D {\bf 32}, 379 (1985).
  doi:10.1103/PhysRevD.32.379


\bibitem{Doneva:2017bvd}
  D.~D.~Doneva and S.~S.~Yazadjiev,
  Phys.\ Rev.\ Lett.\  {\bf 120}, no. 13, 131103 (2018)
  doi:10.1103/PhysRevLett.120.131103
  [arXiv:1711.01187 [gr-qc]].

\bibitem{Silva:2017uqg}
  H.~O.~Silva, J.~Sakstein, L.~Gualtieri, T.~P.~Sotiriou and E.~Berti,
  Phys.\ Rev.\ Lett.\  {\bf 120}, no. 13, 131104 (2018)
  doi:10.1103/PhysRevLett.120.131104
  [arXiv:1711.02080 [gr-qc]].

\bibitem{Herdeiro:2018wub}
  C.~A.~R.~Herdeiro, E.~Radu, N.~Sanchis-Gual and J.~A.~Font,
  Phys.\ Rev.\ Lett.\  {\bf 121}, no. 10, 101102 (2018)
  doi:10.1103/PhysRevLett.121.101102
  [arXiv:1806.05190 [gr-qc]].

\bibitem{Horndeski:1974wa}
  G.~W.~Horndeski,
  Int.\ J.\ Theor.\ Phys.\  {\bf 10}, 363 (1974).
  doi:10.1007/BF01807638

\bibitem{Kobayashi:2014wsa}
  T.~Kobayashi, H.~Motohashi and T.~Suyama,
  Phys.\ Rev.\ D {\bf 89}, no. 8, 084042 (2014)
  doi:10.1103/PhysRevD.89.084042
  [arXiv:1402.6740 [gr-qc]].

\bibitem{Ganguly:2017ort}
  A.~Ganguly, R.~Gannouji, M.~Gonzalez-Espinoza and C.~Pizarro-Moya,
  Class.\ Quant.\ Grav.\  {\bf 35}, no. 14, 145008 (2018)
  doi:10.1088/1361-6382/aac8a0
  [arXiv:1710.07669 [gr-qc]].


\bibitem{Martinez:2002ru}
  C.~Martinez, R.~Troncoso and J.~Zanelli,
  Phys.\ Rev.\ D {\bf 67}, 024008 (2003)
  doi:10.1103/PhysRevD.67.024008
  [hep-th/0205319].

\bibitem{Charmousis:2014zaa}
  C.~Charmousis, T.~Kolyvaris, E.~Papantonopoulos and M.~Tsoukalas,
  JHEP {\bf 1407}, 085 (2014)
  doi:10.1007/JHEP07(2014)085
  [arXiv:1404.1024 [gr-qc]].


\bibitem{Myung:2014nua}
  Y.~S.~Myung and T.~Moon,
  Phys.\ Rev.\ D {\bf 89}, no. 10, 104009 (2014)
  doi:10.1103/PhysRevD.89.104009
  [arXiv:1401.6862 [gr-qc]].


\bibitem{Babichev:2013una}
  E.~Babichev and A.~Fabbri,
  Class.\ Quant.\ Grav.\  {\bf 30}, 152001 (2013)
  doi:10.1088/0264-9381/30/15/152001
  [arXiv:1304.5992 [gr-qc]].



\bibitem{Brito:2013wya}
  R.~Brito, V.~Cardoso and P.~Pani,
  Phys.\ Rev.\ D {\bf 88}, no. 2, 023514 (2013)
  doi:10.1103/PhysRevD.88.023514
  [arXiv:1304.6725 [gr-qc]].


\bibitem{Gregory:1993vy}
  R.~Gregory and R.~Laflamme,
  Phys.\ Rev.\ Lett.\  {\bf 70}, 2837 (1993)
  doi:10.1103/PhysRevLett.70.2837
  [hep-th/9301052].


\bibitem{XZ}
B.~C. Xanthopolos and T.~Zannias, J.\ Math.\ Phys.(N.Y.) {\bf 32}, 1875(1991).

\bibitem{Charmousis:2015aya}
  C.~Charmousis and D.~Iosifidis,
  J.\ Phys.\ Conf.\ Ser.\  {\bf 600}, 012003 (2015)
  doi:10.1088/1742-6596/600/1/012003
  [arXiv:1501.05167 [gr-qc]].


\end{thebibliography}
\end{document}